# DISTRIBUTED VERTEX COVER ALGORITHMS FOR WIRELESS SENSOR NETWORKS

Vedat Kavalci[1], Aybars Ural[2] and Orhan Dagdeviren[2]

[1]School for Vocational Higher Education, Izmir University, Izmir, Turkey
[2]International Computer Institute, Ege University, Izmir, Turkey

## ABSTRACT

*Vertex covering has important applications for wireless sensor networks such as monitoring link failures, facility location, clustering, and data aggregation. In this study, we designed three algorithms for constructing vertex cover in wireless sensor networks. The first algorithm, which is an adaption of the Parnas & Ron's algorithm, is a greedy approach that finds a vertex cover by using the degrees of the nodes. The second algorithm finds a vertex cover from graph matching where Hoepman's weighted matching algorithm is used. The third algorithm firstly forms a breadth-first search tree and then constructs a vertex cover by selecting nodes with predefined levels from breadth-first tree. We show the operation of the designed algorithms, analyze them, and provide the simulation results in the TOSSIM environment. Finally we have implemented, compared and assessed all these approaches. The transmitted message count of the first algorithm is smallest among other algorithms where the third algorithm has turned out to be presenting the best results in vertex cover approximation ratio.*

## KEYWORDS

*Vertex Cover, Wireless Sensor Networks, Distributed Algorithms, Approximation Algorithms, Greedy Algorithms, Graph Matching, Breadth-First Search Tree.*

## 1. INTRODUCTION

Wireless sensor networks (WSNs) are formed by grouping tiny, self-organized, autonomously running, and generally radio-communicable and smart sensor devices into a network in some specific geographical region. Although there may appear differences borne by their usage type and aim, the main common feature of these devices is the limitedness of sources. Basically, these limited characteristics are little physical dimensions, little power sources, short range of radio, little memory capacity, lack of information about the other parts of the network and simplicity of the communication skills [1].

Due to these mentioned limitedness, rather than a centralized fashion algorithmic approach, distributed algorithms in WSN are preferred where each node runs the same code with little or no difference than the others. In WSN, distributed algorithms let nodes to communicate and pass messages carrying data to other nodes and/or to some specific reception points. Distributed algorithms have to be designed in such a way that the nodes have to use the least number of software and hardware components. In this manner, they have to send as little number of messages as possible, the messages have to be as small as possible and the processing of data and messages on the nodes has to be minimized.

A WSN can be modeled with an undirected graph $G=(V,E)$ where $V$ is the set of vertices and $E$ is the set of edges. Graph-theoretical structures can be used for WSNs to solve problems such as building communication infrastructures for reacting topology changes. Vertex cover (VC) is one of these technique and it is quite useful when combined with the power of distributed nature of the WSN. A vertex cover of a graph is the set of vertices in which at least one end of each and every edge of it is included. The elements of a cover set can be used for various purposes, since every communication link will be under the coverage of one or more nodes. Backbone formation, data aggregation management, cluster formation and management, hub or router location designation and traffic control on the information flow are some of these [2]. The problem of finding the minimum vertex cover set (VCS) in a system containing various interconnections which can be modeled as a graph is an NP-complete problem. However, simple approximation algorithms can efficiently find a cover that yields a result set with the size of a multiple of the number of the elements of the minimum cover [3, 4, 5].

In this study, we designed and implement three different distributed vertex cover approximation algorithms and measure the performances of them. The first algorithm is a greedy approach that is adapted from Parnas & Ron's study [6] and the second algorithm uses graph matching [7]. The third algorithm has been designed depending on the idea of a Breadth-First Search (BFS) tree [3, 8]. When forming the BFS tree, the vertex cover itself is formed as well and this constitutes the main improvement. Thus, reductions in processing time and energy saving have been achieved. We first analyze the time, space and message complexities of the algorithms and compare them. After that, we provide practical evaluation by making simulations of the algorithms on TOSSIM simulator.

The rest of this paper is organized as follows: in Section 2, the problem formulation is given. The related work about distributed vertex covering is surveyed in Section 3. The implemented algorithms are described in detail in Section 4. In Section 5, simulation measurements of the algorithms are introduced. Lastly, conclusions are given in Section 6.

## 2. PROBLEM FORMULATION

In this section, we first identify the underlying network model than we define VC problem for WSNs. Our network model is depicted in Figure 1.a and its underlying communication graph in Figure 1.b where following assumptions are made:

- Each node in the network has a unique identifier number.
- The communication links are unidirectional and unweighted.
- Each node is identical in terms of hardware and software.
- Each node has identical transmission range and knows its neighbors.
- Nodes are time-synchronized to execute round-based synchronous algorithms. Time synchronization can be done by applying a method such as [9].

Our objectives for VC construction in sensor networks can be listed as follows:

- VCS should be as small as possible. For example the $VCS_1$ in Figure 1.c is {$A,B,D$} and has 3 elements where the $VCS_2$ in Figure 1.d is {$A,C$} and has 2 elements. Thus, $VCS_2$ is preferable over $VCS_1$.
- The VC algorithm should be independent from the underlying MAC layer.
- The VC algorithm should be efficient in terms of time, space and message complexities in order to provide low energy consumptions.

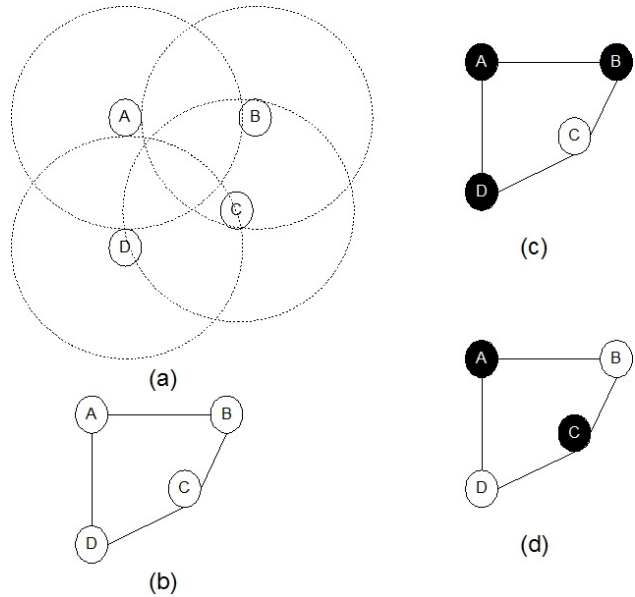

Figure 1. (a) A wireless sensor network (b) The communication graph
(c) The vertex cover VCS$_1$={A,B,D} (d) The vertex cover VCS$_2$={A,C}

# 3. RELATED WORK

In a graph, the number of elements of a minimum VCS may vary between *1* and one less than the number of the vertices depending on the topology. Hence, by *$2^n$-1* trials, VCS possibilities can be evaluated and the minimum VCS can be found [10]. As a result of this identification, such a brute-force algorithm will surely have a time complexity of *O($2^n$)*. With an exponential time complexity algorithm and with a graph of vertices more than 20 in number, to find a solution for the minimum VCS requires a huge processing power and takes unfeasibly long processing time.

Therefore, when such problems are in hand, employing approximation algorithms which present fast and close-to-ideal results is a considerably feasible choice. Approximation algorithms have polynomial time complexities and hence they generate the required results in polynomial time. But these algorithms do not put forward the best results; they approximate the best result which can be considered as an optimal result. By a centralized algorithm, the approximation ratio for a minimum VCS is at least *1.36*. The best upper bound known is $2 - \Theta(1/\sqrt{log_2(n)})$ [11, 12, 13] .

When forming the minimum VCS via matching, maximum matching should be attained where there shouldn't be any vertices that are not matched. This situation is only met when the graph is a bipartite graph. If it is so, then we can use one vertex from each match and form the minimum VCS. But usually many graphs are not bipartite and a maximum matching set will not lead us to a minimum VCS by only taking one vertex from each matching couple. Hence some of the graph will remain uncovered. In such cases, one solution is to take both of the matching vertices into the VCS. But, this might lead to high approximation ratio and diverging from minimum cover.

In this study, we surveyed the former studies which made assumptions similar to ours. In other words, we have surveyed studies which consider WSN conditions, with an unweighted, unidirectional and connected network, and assuming to run on a distributed fashion. The findings are given in Table 1 where Δ is the network degree and *n* is the node count. Distributed VC algorithms generally use three main methods to minimize VCS. The first method is a greedy approach which considers the degree of vertices. The second method aims to find a maximum matching and then forms the VCS from this matching. The last method uses independent set to find VCS.

Table 1. Theoretical performance results for some algorithms [14]

| Approximation Ratio | Time Complexity | Year | Algorithm |
|:---:|:---:|:---:|:---:|
| 2 | $O(\log^4{}_2(n))$ | 1997 | [15](with matching) |
| 2 | $O(\Delta + \log^*{}_2(n))$ | 2001 | [16](with matching) |
| 2 | $O(\Delta^2)$ | 2009 | [17](with edge packing) |
| 3 | $O(\Delta+1)$ | 2009 | [18](with port numbering) |
| 2.5 | $O(\Delta)$ | 2010 | [12]( with partitioning) |

# 4. DESIGNED ALGORITHMS

We designed 3 VC algorithms for WSNs. The first algorithm forms a VC by using a greedy approach. The second algorithm produces a VC from graph matching. The third algorithm forms a BFS tree then constructs VC. The details of the algorithms and performance analyses are presented in the following sections.

## 4.1. Basic Distributed Greedy Vertex Cover Algorithm

This algorithm is an adaptation of Parnas and Ron's algorithm [6] for sensor networks. In this algorithm, each vertex of certain degree adds itself to the VCS and disconnects itself from the graph. We assume that all nodes know Δ value. This operation can be applied by broadcast and convergecast operations. Messaging operations amongst the nodes require acknowledgement procedures. Since the algorithm run in a distributed manner, a general looping system is employed and it is supported via message transmission between the nodes.

We have used the condition for selection of the vertex for the VCS as "*greater than*" rather than "*greater than or equal to*", which takes place in the original form of this algorithm; in order to get closer results to our target obtaining a VCS with the minimum number of elements. The pseudocode of the algorithm is given in Algorithm 1.

```
BOOT:
     Just after boot, every node calculates its own degree in the graph (MyDegree)
     R=0 (rounds' counter is set to zero)
     T2 (new round timer) is set for one shot
     T0 (periodic timer for sending drop messages) is set as periodical

UPON T0 FIRED:
    If there exists messages in queue and an ACK is not awaited,
         Send one message from the queue
         Do the actions for the ACK of the sent message and set T1 to one shot for ACK

UPON T1 FIRED:
    Resend the message for which an ACK is not received and set T1 to one shot for ACK
```

```
        again

UPON SENDDONE:
    Message is ensured that it has been sent
    Stop T1
    Set AckWait = False

UPON T2 FIRED:
    If ACK is not awaited and there exists no messages in the queue

        R = R+1 (increment rounds' counter by one)
        If R > log₂(Δ)  then return (termination condition is met)
        If (MyDegree) > (Δ / (2ᴿ)) then
            Advertise ownself that it is in the cover set
            Send "drop" message to all non-zero degree neighbors (once it is within the cover set,
                it no longer needs to set timer T2)
        Else set T2 to one shot

UPON RECEIVE:
    If "drop me" message is received,
        Delete the appropriate node from the adjacency list
        MyDegree  = MyDegree -1
```

Algorithm 1. Pseudocode of the Basic Distributed Greedy Vertex Cover Algorithm

## 4.2. Distributed Vertex Cover via Greedy Matching Algorithm

Our second algorithm benefits from graph matching to construct a VCS. We use Hoepman's weighted graph matching algorithm [7] then add each endpoint of the matching set to a VCS. Also, the algorithm in hand uses the notion of weights assigned to the edges in the graph. But contrarily, our aim is to approximate a distributed vertex cover in an unweighted graph, so we do not use edge weights.

Our implementation is realized through certain assumptions as well. All messaging will be finished while timer T2 is running; therefore it is set to quite a high value. All messaging operations amongst the nodes require acknowledgement procedures. A general looping system is employed and it is supported via message transmissions. The pseudocode of the distributed vertex cover via greedy matching is given in Algorithm 2.

```
BOOT:
    At the beginning each node sends its degree to the other nodes (This takes place at round
        zero)
    Timers T2, T0, T1 are set; value of T2 quite high.

UPON T0 FIRED:
    If there is a message in the queue and an ACK is not awaited,
        Send one message from the queue
        Do the actions for the ACK of the sent message and set T1 to one shot for ACK

UPON T1 FIRED:
    Resend the message for which an ACK is not received set T1 to one shot for ACK again

UPON SENDDONE:
```

```
    Message is ensured that it has been sent
    Stop T1
    Set AckWait = False

UPON T2 FIRED:
    If ACK is not awaited and there exists no messages in the queue, (normally the queue is
        expected to be empty since the messaging time match request is designated to be long)

        Sort the active adjacency list by the degree of the nodes in descending order
        R = R+1 (increment rounds' counter by one)
        Send match request to the suitable neighbor (who is the neighbor with the least degree –
            that is the last one in the list which has been just sorted)

UPON RECEIVE:
    If a "request" message is received
        If the requester is also requested by me, (I have found my match)
            Stop timer T2
            Fulfill the following matching operations:
                Zero own degree and inform all neighbors about this (send a "drop" message)
                    If round ==1
                        The node in the matching couple who has the higher degree places itself in
                            the vertex cover set
                        If the original degrees of the nodes in the matching couple are equal,
                            The node with the higher ID number places itself in the vertex cover set
                    If round >1
                        Both of the nodes in the matching couple place themselves in the vertex
                            cover set

    If a "degree" message is received
        Update the degree of the sender node in the adjacency list
        If the received "degree" message includes zero degree, (this in fact means a "drop"
            message)
            Delete this node from the neighbor list
            Inform all neighbors about the change by means of a "drop" message
```

Algorithm 2. Pseudocode of the Distributed Vertex Cover via Greedy Matching Algorithm

## 4.3. Distributed Vertex Cover Algorithm with Breadth-First Search Tree

Our third algorithm builds a BFS tree and constructs a VCS by using this infrastructure. BFS tree formation is very important for WSNs since BFS provides a routing backbone. In a BFS tree, each node is associated with a level value which is its hop distance from the root (sink) node.
In our algorithm, each node will advertise its level information to its neighbors. Each node which has once got the level information of its neighbors, will decide, through certain rules (as given below), whether it should participate within the set of the vertex covering nodes or not. If the level value of a node is even, then it will be directly in the set. If the level value of a node is odd and it has a neighbor which has also an odd level value, then the node which has the larger ID number in-between these two will decide that if it is in the VCS or not. The algorithm is given in Algorithm 3.

```
BOOT:
    Sink node:
        It forms its own neighbor list using the adjacency list.
        It sends "level 0" infrastructure message to all its neighbors.
```

Timer T0 is set to periodic for the messages.
Timer T2 is set for one shot to be used in the discovering "parents" and "levels".

Other nodes:
They form their own neighbor lists using the adjacency lists.
Timer T0 is set to periodic for the messages.
Timer T2 is set for one shot to be used in the discovering "parents" and "levels".

UPON T0 FIRED:
If there is a message in the queue and an ACK is not awaited,
Send one message from the queue.
Do the actions for the ACK of the sent message and set T1 to one shot for ACK.

UPON T1 FIRED:
Resend the message for which an ACK is not received.
Set T1 to one shot for ACK again.

UPON SENDDONE:
*Message is ensured that it has been sent.*
Stop T1.
Set AckWait = False.

UPON T2 FIRED:
*The BFS tree is constructed.*
*Each node now knows its "parent" and its "level" value.*
Timer T3 is set for one shot in order all nodes to decide whether they are within the vertex
cover set or not.
All nodes send their "level" information to all their neighbors.

UPON T3 FIRED:
*Every node is ensured that it has got information about each of its neighbors.*
If own "level" is even then place ownself in the vertex cover set.
If own "level" is odd then,
Search for a neighbor whose "level" is also odd
If found then place the node with the higher ID in the vertex cover set.

UPON RECEIVE:
If an "infrastructure" message is received
If MyLevel > MsgSenderLevel +1 *(if a lower level parent is found)*
Set MyLevel = MsgSenderLevel +1
Set MyParent = MsgSenderId
Send "level" information to all neighbors but MsgSender

If a "level" message is received
Update own neighbor list depending on the MsgSenderId

Algorithm 3. Pseudocode of the Distributed Vertex Cover with Breadth-First Search Algorithm

# 5. PERFORMANCE EVALUATIONS

In this section, we provide theoretical and practical evaluations of the algorithms.

## 5.1. Theoretical Evaluation

We have performed the analysis through time, message and space complexity calculations. We have evaluated the worst case performances of the algorithms. Space complexity is given per node while time and message complexities are calculated network wide. Theorems 1 to 9 are given for the theoretical analysis.

**Theorem 1.** *The time complexity of the basic distributed greedy algorithm is $O(log_2(n))$.*

*Proof.* The worst case happens when at each round only one node advertises itself to be in the vertex cover set (VCS). In such a case, the maximum number of rounds is reached. Since each node which enters the VCS sends a drop message only to its neighbors, the messaging time is *1* unit. The maximum number of rounds possible for this algorithm is $R = log_2(\Delta)$ , where *Δ* is the maximum degree of the graph. Δ can take *n-1* as maximum value. In this case, maximum number of messages become $O(R)$. Hence the time complexity becomes $O(log_2(n-1)) = O(log_2(n))$.

**Theorem 2.** *The message complexity of the basic distributed greedy algorithm is $O(n^2)$.*

*Proof.* A node can send at most as many number of messages as the number of its neighbors. Hence message complexity is Δ per node. The worst case can arise in case of a fully connected graph where Δ=n-1 and for all the *n* nodes the overall message complexity becomes $O(n*(n-1)) = O(n^2)$.

**Theorem 3.** *The space complexity of the basic distributed greedy algorithm is $O(n)$ per node.*

*Proof.* Each node has to use memory for keeping its own neighbor list and the maximum degree *Δ* present in the graph. If the graph is a fully connected one, the neighbor list can have at most *n-1* elements. Hence, the space complexity of this algorithm in the worst case is $O(n-1+\Delta) = O(n)$.

**Theorem 4.** *The time complexity of the distributed cover via matching algorithm is $O(n)$.*

*Proof.* In the beginning, each node sends its degree to its neighbors, sends a matching request to one neighbor and may send a drop message to its neighbors. The time complexity of these operations are constant. In the worst case, maximum number of rounds appears when only two nodes match in every round. In such a case, number of rounds is *|V|/2=n/2*. Hence, time complexity becomes $R=n/2$. Conclusively, $O(n)$ is the worst case complexity.

**Theorem 5.** *The message complexity of the distributed cover via matching algorithm is $O(n)$.*

*Proof.* Initially, each node sends its degree to all of its neighbors. This adds up to *d* messages, *d* being the degree of the node. In worst case, if the graph is fully connected, $\sum\Theta(d) = O(|E|) = O(n^2)$. In every round, request and drop messages are sent. Total request messages are $\sum\Theta(d) = O(|E|) = O(n^2)$. Similarly, the complexity of drop messages is $O(|E|) = O(n^2)$. Hence, overall message complexity is $O(n^2)$.

**Theorem 6.** *The space complexity of the distributed cover via matching algorithm is $O(n)$ per node.*

*Proof.* Each node has to use memory for keeping both for keeping its neighbors' original degrees and their changing degrees within the rounds. This means a memory space two times the message count. And a memory space is used for keeping the ID of the matched node and another memory space is used for storing the round info. Thus space complexity is $\Theta(2*d+c) = \Theta(2*(n-1)+c)$. Degree $d$ may be at most $n-1$, hence in the worst case space complexity per node is $O(n)$.

**Theorem 7.** *The time complexity of the distributed cover algorithm with BFS tree is $O(n)$.*

*Proof.* At the beginning, BFS tree is constructed in $O(D)$ times where $D$ is the diameter of the graph. Afterwards, each node will send to its neighbors its own level value. The cost of this operation is $O(1)$, and if the BFS tree time cost is added to this, it sums up to $O(n)+O(1)=O(n)$. Total number of rounds passed during the execution of this algorithm is just two, and this value is independent of the number of nodes present in the graph. In the first round, the BFS tree is constructed. This procedure takes place in an asynchronous manner. In the second round, level values are shared and the VCS is constructed.

**Theorem 8.** *The message complexity of the distributed cover algorithm with BFS tree is $O(n^3)$.*

*Proof.* Initially, BFS tree construction has a cost of $O(n^3)$. Afterwards, each node will send to its neighbors its own level value. The overall number of neighbors of an $n$ nodes graph is $n*(n-1)$. So, in the worst case the messaging cost will yield into $\sum \Theta(d)=O(|E|)=O(n^2)$. The total message cost is $O(n^2)+O(n^3)=O(n^3)$.

**Theorem 9.** *The space complexity of the distributed cover algorithm with BFS tree is $O(n)$ per node.*

*Proof.* Each node has to keep the tree level information of its neighbor nodes. Besides, it has to use a memory space for saving its own tree level information. Hence, $O(d+c)$ memory is required per node. When the overall memory cost is calculated with the memory cost of BFS tree construction, it sums up to $O(d+2c)$. Since $d$ can be at most $n-1$ in an n nodes graph, overall space complexity becomes $O(n-1+2c)=O(n)$.

Table 2. Overall theoretical results for worst cases

| Algorithm | Time Complexity | Message Complexity | Space Complexity |
|---|---|---|---|
| Basic Distributed Greedy Vertex Cover Algorithm | $O(log_2(n))$ | $O(n^2)$ | $O(n)$ |
| Distributed Vertex Cover via Greedy Matching Algorithm | $O(n)$ | $O(n^2)$ | $O(n)$ |
| Distributed Vertex Cover Algorithm with BFS Tree | $O(n)$ | $O(n^3)$ | $O(n)$ |

When we inspect the theoretical performances, we find out that the best one is the first algorithm. We have observed the worst performance with the BFS tree based algorithm. The reason is asynchronous BFS construction is a costly operation. Most of the message complexity stems from the infrastructure build-up portion, and hence this dominates the overall message complexity. If a more effective way of tree formation than the BFS can be discovered, the message complexity can be reduced. The results tabulated in Table 2 represent the worst case scenarios. Since these results are not tight bounds, they only indicate the order of magnitude for the complexities. And also they do not give the complete idea for the effectiveness of the algorithm for finding a minimal VCS. To fulfill our analysis, we present simulation measurements for the practical evaluation in the following section.

## 5.2 Practical Evaluation

We have implemented and tested the designed algorithms on TOSSIM (TinyOS Simulator) [19] using nesC (networked embedded systems C) language [20]. This language is developed for writing applications special to embedded systems where its features like event driven structure, flexible concurrency model and component oriented application design are very useful for such simulation purposes. It runs in full harmony with the simulation environment TOSSIM.

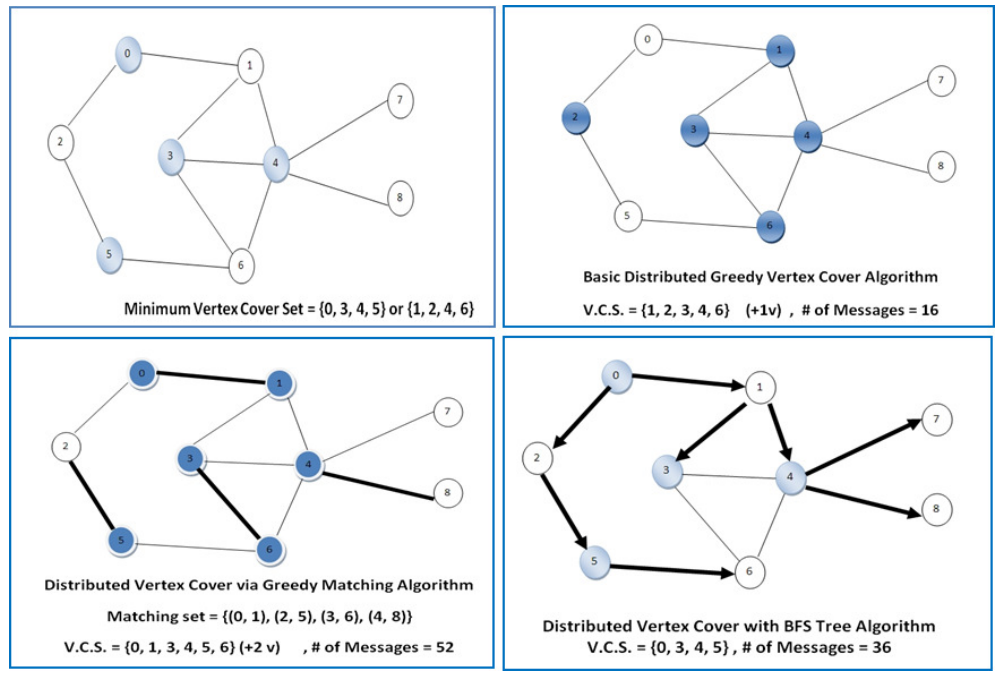

Figure 2. VCSs on the Basic Graph

Firstly, we measured approximation ratios of the algorithms. Then, we assessed the simulation results by comparing them with the optimum results. For achieving this goal, we developed a brute-force vertex cover set finder algorithm. We have also taken into account the number of messages which each algorithm yields. We compared the number of messages with respect to each other for various topologies. Since, we have made the assumption that the round duration for all of the algorithms will be long enough to let all the messaging, runtime measurements are not meaningful. Therefore, we have not measured the durations and hence have not taken then into our assessment.

We have run the simulations on four different graphs. First one is a simple graph designed for general testing purposes. The second one is called octopus and designed especially for testing the pitfalls that the algorithms may face. The third one is the famous and well-known Hamiltonian graph in which every node has equal degrees. The last one is a randomly generated graph and lets us to observe the effects of increasing node number on the algorithms. The measurements for the simple graph, octopus and Hamiltonian graph are given in Figures 2, 3 and 4 respectively. For all graphs, basic algorithm (first algorithm) has the best transfer performance. For simple graph and Hamiltonian graph BFS based algorithm (third algorithm) has the best performance in terms of approximation ratio. For octopus graph, basic algorithm has the best approximation ratio performance ratio.

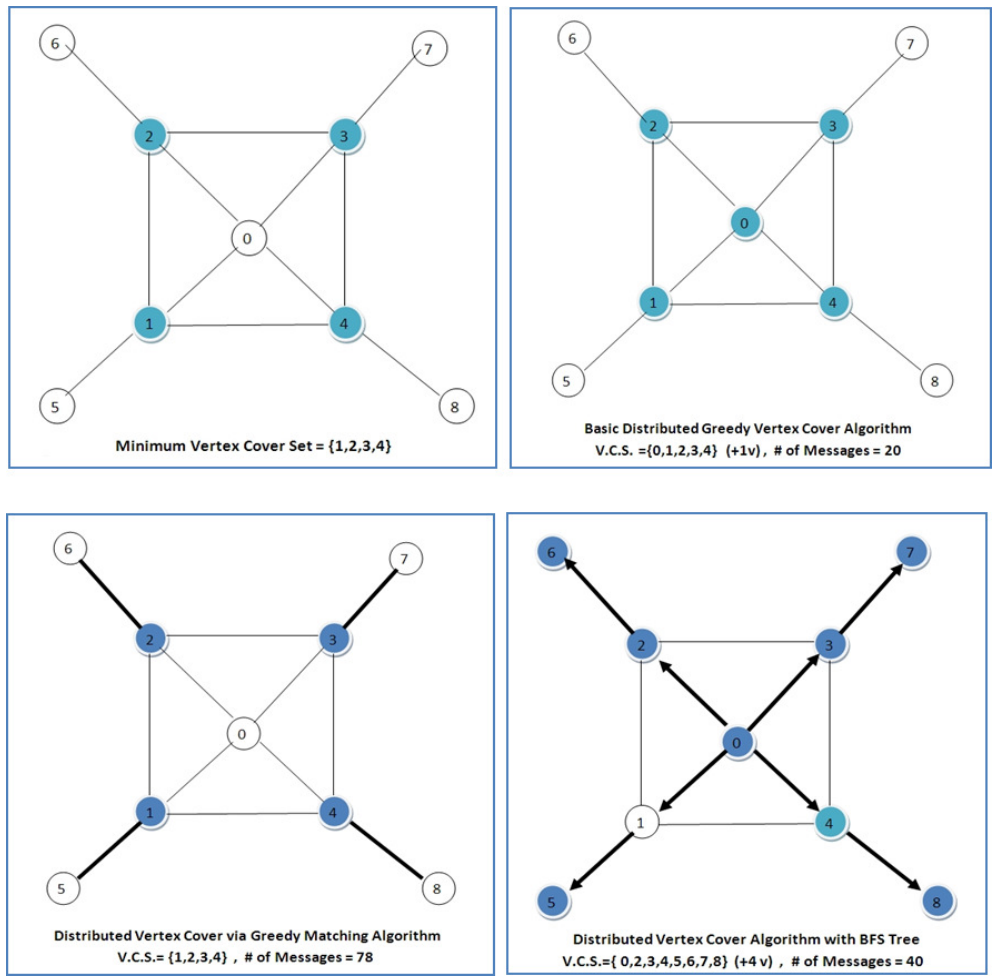

Figure 3. VCSs on the Octopus Graph

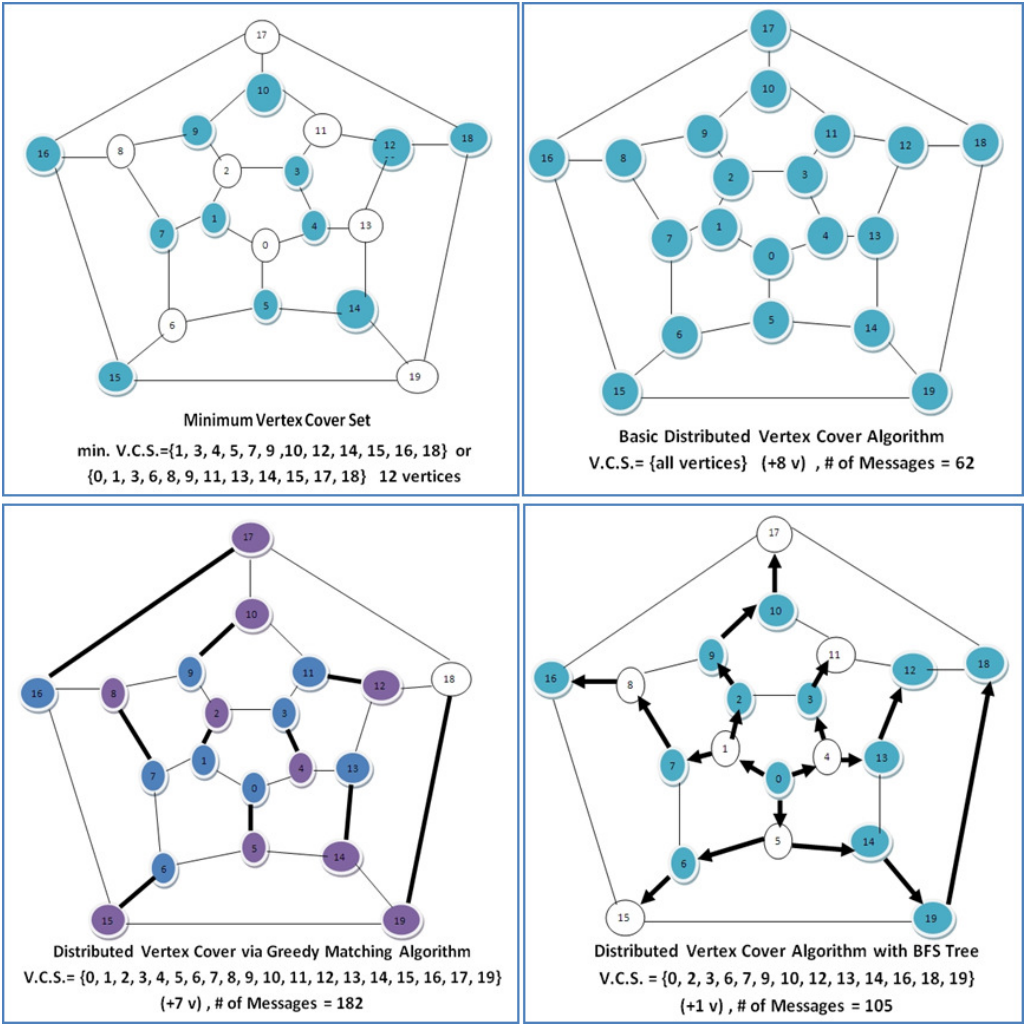

Figure 4. VCSs on the Hamiltonian Graph

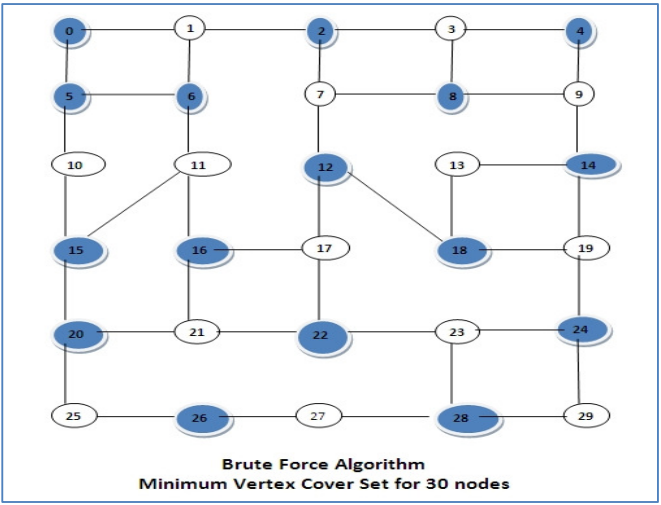

Figure 5. Minimum VCS for the test graph found by brute-force

The graph in Figure 5 is formed in a randomized way so that it can be a small size representation of a real-life WSN model. We have numbered the vertices from 0 to 29 and for our tests we used the first 10, the first 20 and then all of the vertices. The simulation results of the algorithm are given in Table 3.

Table 3. Overall practical results

| GRAPH-1 | BASIC GRAPH | | | |
|---|---|---|---|---|
| | Number of Elements of Vertex Cover Set | Number of Messages | Results | |
| | | | Deflection from min. V.C.S. | Approximation Ratio |
| Brute Force | 4 | N/A | 0 | 1 |
| Alg. No.1 | 5 | 16 | 1 | 1,25 |
| Alg. No.2 | 6 | 52 | 2 | 1,5 |
| Alg. No.3 | 4 | 36 | 0 | 1 |

| GRAPH-2 | OCTOPUS GRAPH | | | |
|---|---|---|---|---|
| | Number of Elements of Vertex Cover Set | Number of Messages | Results | |
| | | | Deflection from min. V.C.S. | Approximation Ratio |
| Brute Force | 4 | N/A | 0 | 1 |
| Alg. No.1 | 5 | 20 | 1 | 1,25 |
| Alg. No.2 | 4 | 78 | 0 | 1 |
| Alg. No.3 | 8 | 40 | 4 | 2 |

| GRAPH-3 | HAMILTONIAN GRAPH | | | |
|---|---|---|---|---|
| | Number of Elements of Vertex Cover Set | Number of Messages | Results | |
| | | | Deflection from min. V.C.S. | Approximation Ratio |
| Brute Force | 12 | N/A | 0 | 1 |
| Alg. No.1 | 20 | 62 | 8 | 1,667 |
| Alg. No.2 | 19 | 182 | 7 | 1,583 |
| Alg. No.3 | 13 | 105 | 1 | 1,083 |

| GRAPH -4 | TEST GRAPH | | | | | | | | | | | |
|---|---|---|---|---|---|---|---|---|---|---|---|---|
| | Number of Elements of Vertex Cover Set | | | Number of Messages | | | Results | | | | | |
| | 10 Nodes | 20 Nodes | 30 Nodes | 10 Nodes | 20 Nodes | 30 Nodes | 10 Nodes | | 20 Nodes | | 30 Nodes | |
| | | | | | | | Deflection | Ratio | Deflection | Ratio | Deflection | Ratio |
| Brute Force | 5 | 10 | 16 | N/A | N/A | N/A | 0 | 1 | 0 | 1 | 0 | 1 |
| Alg. No.1 | 10 | 20 | 30 | 24 | 52 | 86 | 5 | 2 | 10 | 2 | 14 | 1,875 |
| Alg. No.2 | 8 | 16 | 25 | 60 | 131 | 216 | 3 | 1,6 | 6 | 1,6 | 9 | 1,563 |
| Alg. No.3 | 5 | 12 | 18 | 40 | 89 | 141 | 0 | 1 | 2 | 1,2 | 2 | 1,125 |

**Alg. No.1** : Basic Distributed Greedy Vertex Cover Algorithm
**Alg. No.2** : Distributed Vertex Cover via Greedy Matching Algorithm
**Alg. No.3** : Distributed Vertex Cover Algorithm with BFS Tree

The basic and matching based algorithms divert from the optimum results while the node number is increasing as seen in Table 3. On the contrary, BFS tree based algorithm converges to the optimum result meanwhile. That is, the approximation ratios of the basic and matching based algorithm worsen while BFS tree based algorithm gets better. BFS tree based algorithm yields worse results than the other two in the octopus graph. The reason of this fact is BFS tree based algorithm performs worse for dense networks.

Although the basic algorithm has the least number of messages, its approximation to the optimum VCS value yields the worst value. The matching based algorithm has the worst message transfer performance but it is better than basic algorithm in terms of approximation ratio. BFS tree based algorithm is worse message transfer performance than the basic algorithm but it yields best VC approximation as also shown in Figure 6. At this point, it should be regarded that the BFS tree based algorithm serves to establish an infrastructure that maintains two different purposes at the same time.

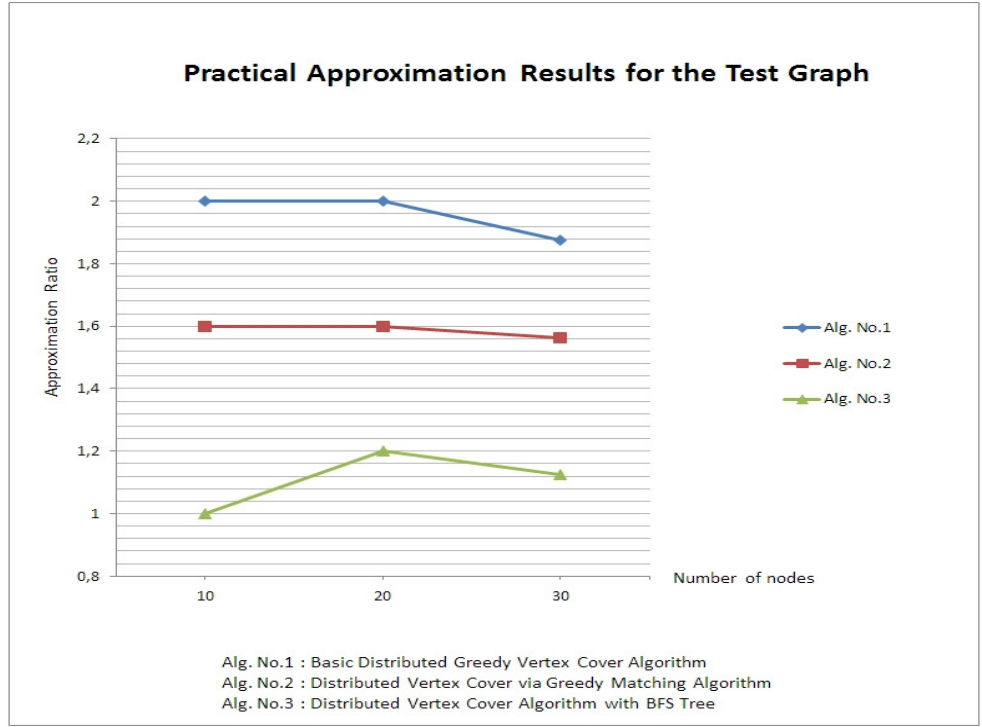

Figure 6. Comparison of VCS approximation ratio results for the test graph

# 6. CONCLUSIONS

In this paper, we designed three distributed VC algorithms for WSNs to be used as a communication infrastructure. The first algorithm is a greedy approach which is WSN adapted version of Parnas & Ron's algorithm and used vertex degrees. The second algorithm finds VC from graph matching where Hoepman's weighted graph matching algorithm is adapted. The third algorithm is a BFS tree based approach in which a routing infrastructure is constructed with a VC. The algorithms are theoretically analyzed and simulated in TOSSIM simulation environment.

Although the "*Basic Distributed Greedy Vertex Cover Algorithm*" results in the least number of messages amongst the three, it gives out the worst approximation ratio. The "*Distributed Vertex Cover via Greedy Matching Algorithm*" presents a message transfer performance in-between the other two and it yields an approximation ratio in-between those of the other two as well. The "*Distributed Vertex Cover Algorithm with BFS Tree*" produces a results for the number of messages in-between those of the other two while it issues the best approximation ratio. And yet an additional feature of the third algorithm besides these results is its functionality in building an effective communication infrastructure between the sink node and the leaf nodes. These results show us that the designed VC algorithms are good candidates for infrastructure management in WSNs. A subject for the future studies can be the design distributed algorithms for other distributed systems such as mobile ad hoc networks [21] and grids [22].

**Authors**

**Vedat Kavalci**

Vedat Kavalcı received the BSc. degree in Computer Eng. and MSc. degree in Computer Eng. from Ege University. He is a lecturer in Vocational Higher School in Izmir University. He is currently a Ph.D. student in Computer Science and Information Technology joint program at Ege and Izmir Universities. His interests lie in the computer networking, wireless sensor networks and embedded systems areas.

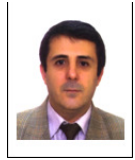

**Aybars Ural**

Aybars Ural received one of his BSc. degrees in Mechanical Eng. from Middle East Technical University and the other one in Computer Eng. from Yasar University. He is currently a Ph.D. student in Information Technology program at International Computing Institute in Ege University. His research interests are mobile informatics and decision making software.

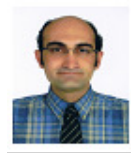

**Orhan Dagdeviren**

Orhan Dagdeviren received the BSc. degree in Computer Eng. and MSc. degree in Computer Eng. from Izmir Institute of Technology. He received Ph.D. degree from Ege University, International Computing Institute. He is an assistant professor in International Computing Institute in Ege University. His interests lie in the computer networking and distributed systems areas. His recent focus is on graph theoretic middleware protocol design for wireless sensor networks, mobile ad hoc networks and grid computing.

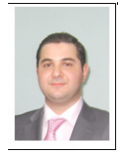